\documentclass[conference]{IEEEtran}
\IEEEoverridecommandlockouts

\usepackage{amsmath,graphicx}
\usepackage{comment,psfrag,amssymb,booktabs,tabularx,pgfplots,color,multirow,rotating,textcomp} % AE

\definecolor{brownish}{rgb}{0.57,0.44,0.44}

\begin{document}

\title{Disparity Estimation for Fisheye Images with an Application to Intermediate View Synthesis}

\author{\IEEEauthorblockN{Andrea Eichenseer, Michel B\"atz, and Andr\'e Kaup}
\thanks{\hspace{-0.275cm}\normalsize \textit{978-1-5090-3649-3/17/\$31.00 \textcopyright2017 IEEE}}
\IEEEauthorblockA{Multimedia Communications and Signal Processing\\
	Friedrich-Alexander University Erlangen-N\"urnberg (FAU)\\Cauerstr. 7, 91058 Erlangen, Germany\\
	Email: \{andrea.eichenseer, michel.baetz, andre.kaup\}@fau.de}}


\maketitle

\begin{abstract}
To obtain depth information from a stereo camera setup, a common way is to conduct disparity estimation between the two views; the disparity map thus generated may then also be used to synthesize arbitrary intermediate views.
A straightforward approach to disparity estimation is block matching, which performs well with perspective data.
When dealing with non-perspective imagery such as obtained from ultra wide-angle fisheye cameras, however, block matching meets its limits.
In this paper, an adapted disparity estimation approach for fisheye images is introduced. 
The proposed method exploits knowledge about the fisheye projection function to transform the fisheye coordinate grid to a corresponding perspective mesh.
Offsets between views can thus be determined more accurately, resulting in more reliable disparity maps. 
By re-projecting the perspective mesh to the fisheye domain, the original fisheye field of view is retained.
The benefit of the proposed method is demonstrated in the context of intermediate view synthesis, for which both objectively evaluated as well as visually convincing results are provided.
\end{abstract}


% \fi
\IEEEpeerreviewmaketitle

%\vspace{0.2cm}
\section{Introduction}
\label{sec:intro}
%\vspace{-0.2cm}

In video surveillance and automotive systems, fisheye cameras~\cite{miyamoto1964fel} that possess an ultra wide field of view (FOV) of 180 degrees and beyond are often employed.
With such an FOV, a single camera is able to capture the entire hemisphere it is facing and thus proves advantageous in surveillance systems~\cite{surveillance}, where tracking~\cite{surveillance2} and detection is important, or in automotive systems to aid the driver~\cite{auto1, auto2, gehrig}.
Contrary to pinhole cameras, the captured images from fisheye cameras exhibit strong radial distortions that cause straight lines of the scene to not be straight in the final images.
While a single fisheye camera already has its advantages, using more cameras offers even more possibilities for detection~\cite{fisheyerobot, pedestrian, fisheyestereo2} and other applications. 
Surround views can be synthesized from multiple fisheye cameras~\cite{liu2008birdseyeview, surroundview}, for instance.
Fisheye stereo is commonly dealt with in the field of computer vision, where abundant literature exists on the topic, e.\,g.,~\cite{gehrig2, fisheyestereo, trinocular}, to name only a few publications. 

\begin{figure}[t]
%\small
\centering
\psfrag{a1}[cr][tc]{{\color{brownish}Left view}}
\psfrag{a2}[cl][tc]{{\color{brownish}Right view}}
\psfrag{a3}[cc][tc]{{\color{brownish}Intermediate view}}
\centerline{\includegraphics[width=0.95\columnwidth]{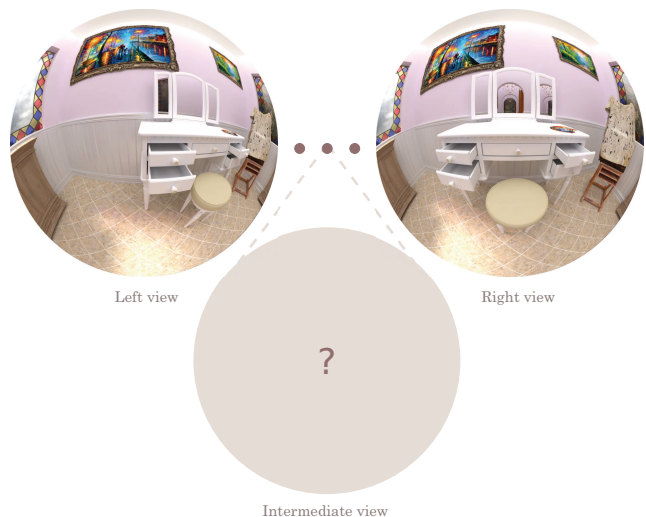}}
\vspace{-0.1cm}
\caption{Fisheye views as obtained by a stereo camera setup. Using disparity estimation, arbitrary intermediate views can be synthesized.}
\label{fig:motivation}
\vspace{-0.2cm}
\end{figure}

A common task in the context of stereo camera setups is the extraction of depth by estimating the disparity between the two available views.
Replacing conventional pinhole cameras with fisheye cameras, new problems arise during disparity estimation as the image characteristics differ significantly due to the non-perspective projection function.
Fig.~\ref{fig:motivation} shows two views, each obtained from a fisheye camera.
The most obvious observation is that the expected horizontal offset between the two views is no longer just horizontal.
While the displacement of the cameras is indeed purely horizontal, the same cannot be said for the image content. The radially distorted objects also change shape as well as vertical position between views, which complicates stereo matching and depth map generation.

In this paper, we investigate block matching for fisheye disparity estimation and adapt it so as to take into account the projection function of the fisheye lens.
The proposed disparity estimation builds upon our motion estimation method for fisheye video sequences~\cite{eichenseer2015motionfish, eichenseer2016motioncalib}, where we could show that incorporating knowledge about the fisheye properties considerably improves the estimation results.
We demonstrate that our adapted disparity estimation method is able to produce more accurate disparity maps compared to the conventional approach by synthesizing intermediate views, for which improved visual results are achieved. An objective evaluation against ground truth data substantiates the visual examples.
A similar approach was introduced in~\cite{vision1}, where the plane-sweeping algorithm was adapted for fisheye stereo. In contrast to that, we focus on a simplistic, versatile approach that finds use in many signal processing and video coding applications.

The remainder of this paper is structured as follows.
Section~\ref{sec:disparity} briefly describes disparity estimation based on block matching and how we employ it. In Section~\ref{sec:fisheye}, we introduce the proposed disparity estimation via fisheye-adapted block matching. Section~\ref{sec:simulation} provides the simulation setup and results, while Section~\ref{sec:conclusion} concludes the paper.

\section{Disparity Estimation via Horizontal Block Matching}
\label{sec:disparity}
%\vspace{-0.2cm}

For setups with conventional cameras that more or less follow the pinhole model, a straightforward approach to estimate the disparity between two views is block matching, which also finds wide use in motion estimation and temporal prediction applications as well as in hybrid video coding.
Assuming rectified views, the block matching can even be restricted to one dimension by making use of epipolar geometry~\cite{hartley2003mvgeo} (e.\,g., the horizontal dimension in case of left and right views) and to one direction (from left to right, for instance).
This results in much fewer candidate blocks to be tested when compared to traditional two-dimensional block matching.

Given two views of a scene, typically a left view $I_\text{left}(m,n)$ and a right view $I_\text{right}(m,n)$, with $m$ and $n$ being the vertical and horizontal spatial coordinates, respectively, the disparity between the two images can be used to estimate one view from the other:
%\vspace{-0.2cm}
\begin{equation}
\tilde I_\text{right}(m,n) = I_\text{left}(m,n+d)\: ,
%\vspace{-0.2cm}
\end{equation}
where $\tilde I_\text{right}(m,n)$ describes the estimated right view and $d$ corresponds to the disparity map entry at position $(m,n)$.
The disparity map itself is denoted as $D(m,n)$ and has the same dimensions as the two views.
In this paper, the disparity map is given from the right view to the left view.

\begin{figure}[t]
%\small
\vspace{0.1cm}
\centering
\psfrag{a1}[tc][tc]{{\color{brownish}Left view}}
\psfrag{a2}[tc][tc]{{\color{brownish}Right view}}
\psfrag{a3}[tc][tc]{$s$}
\psfrag{a4}[tc][tc]{$w$}
\psfrag{a5}[tc][tc]{$b$}
\psfrag{a6}[bc][bc]{$d$}
\psfrag{a7}[bc][bc]{$n$}
\psfrag{a8}[br][br]{$m$}
\centerline{\includegraphics[width=0.89\columnwidth]{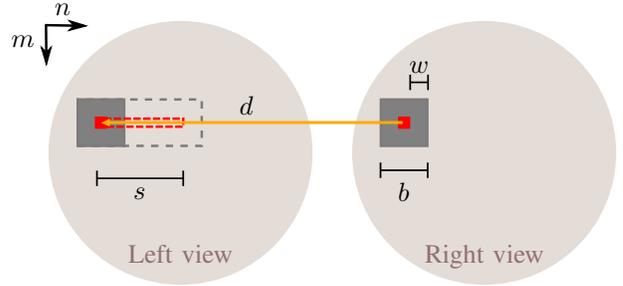}}
\vspace{-0.2cm}
\caption{Horizontal disparity estimation using a support block of size $b\times b$ to determine the disparity $d$ for the red pixel at position $(m,n)$. The area covered by the candidate blocks to be tested is shown by the dashed lines.}
\label{fig:sota}
\vspace{-0.3cm}
\end{figure}

To determine the disparity for each pixel $(m,n)$, we employ block matching in a pixel-wise manner, i.\,e., for each pixel, a support block is defined around the pixel, which is then used for the disparity estimation.
We call the margin around one pixel the support width $w$ in this paper.
The relation between support width $w$ and block size $b\times b$ is given by $b = 2w+1$.
The search range $s$ denotes the maximum offset to be tested during disparity estimation and thus comprises the horizontal offsets in the range of $[0,s]$.
Fig.~\ref{fig:sota} provides a visualization.
For each candidate block within the search range, the current block to be estimated in the right view is compared with the corresponding block in the reference (left) view, shifted by the horizontal offset candidate.
Using the sum of squared differences (SSD) for its simplicity, the error between these two blocks is minimized to find the optimum candidate that describes the disparity between the blocks.
The optimum candidate thus obtained is then only stored for the center pixel of the block and corresponds to the final entry $d$ in $D(m,n)$ at position $(m,n)$.
The disparity $d$ thus describes the offset between the block to be matched in the right view and the best match found in the left view.
Repeating this process for each pixel creates a dense disparity map.

%\vspace{-0.6cm}
%\pagebreak
%\textit{Intermediate View Synthesis}
%\vspace{0.2cm}
\subsection*{Intermediate View Synthesis}

With the disparity from the right view to the left view available, an arbitrary intermediate view can be generated.
In the simplest case, the view that would result from a camera placed exactly in the middle between the two actual cameras is synthesized.
The disparity from the right view to the intermediate view is then exactly half the previously estimated disparity and the intermediate view can thus be calculated by a pixel-wise horizontal shift of the right view:
%\vspace{-0.1cm}
\begin{equation}
%\label{eq:ivs}
I_\text{intermediate}\left(m,n+\frac{1}{2}d\right) = I_\text{right}(m,n)\: ,
%\vspace{-0.1cm}
\end{equation}
followed by a hole-filling algorithm.
Please note that this simple approach can be easily extended towards using both views for the calculation of the intermediate view. To that end, a second disparity map describing the pixel-wise offsets from the left view to the right view would also be necessary.
This is not considered in the scope of this paper, however.

\vspace{0.2cm}
\section{Disparity Estimation via Fisheye-Adapted Block Matching}
\label{sec:fisheye}
\vspace{0.2cm}

Disparity estimation via block matching works well for perspective images.
For fisheye images, however, the horizontal offset is not sufficient to describe the disparity (cf. Fig.~\ref{fig:motivation}).
Also storing the vertical component would result in two disparity maps or a pixel-wise motion vector field containing both horizontal and vertical offsets.
To retain the representation of the disparity by means of a single disparity map so as to also save bits for storage and transmission, and, more importantly, to generate more accurate disparity maps, we therefore propose a novel adaptation to the previously described disparity estimation that takes into consideration the fisheye projection function. We again assume that the two views are rectified, i.\,e., in the perspective representations of the two views, two corresponding points have identical vertical coordinates and the optical axes of the two cameras are parallel.

For our proposed method, we build upon the overlapping block-matching method described in Section~\ref{sec:disparity}.
The one-dimensional block search is not conducted directly in the fisheye image, however, but on projected pixel coordinates. The adaptation described in the following is a novel variant of the fisheye motion estimation method described in~\cite{eichenseer2015motionfish}, that no longer requires a costly hybridization.
With the image center serving as the origin, the pixel positions of the fisheye image are represented in polar coordinates $(r_\text{f},\phi_\text{f})$.
The fisheye projection function is described by a function $p(.)$ that relates the incident angle of light $\theta$ (measured against the optical axis) to the resulting position on the image plane $r_\text{f}$ (measured against the image center): $r_\text{f} = p(\theta)$.
This position $r_\text{f}$ corresponds to the radius of the polar representation. 
A common fisheye model, which we make use of here, is the equisolid angle fisheye projection function given by:
%\vspace{-0.2cm}
\begin{equation}
\label{eq:equisolid}
r_\text{f} = p(\theta) = 2f\sin(\theta /2)\: ,
%\vspace{-0.2cm}
\end{equation}
where $f$ denotes the focal length of the fisheye lens.
Solving~(\ref{eq:equisolid}) for $\theta$ and putting it into the pinhole model $r_\text{p} = f\tan(\theta)$ then gives the projection to perspective coordinates:
%\vspace{-0.2cm}
\begin{equation}
\label{eq:bw}
r_\text{p} = f\tan\left(2 \arcsin\left(\frac{r_\text{f}}{2f}\right)\right)\quad \text{and}\quad\: \phi_\text{p} = \phi_\text{f} \: .
%\vspace{-0.1cm}
\end{equation}
Note that this operation is solely based on pixel positions and as such, only the coordinates are manipulated. This means that no actual distortion correction is performed, which would require interpolation of the luminance values and also result in a significant loss of image content.
This is true for all steps of the adaptation so that all information available in the fisheye views is retained.
In the next step, the obtained perspective polar coordinates $(r_\text{p},\phi_\text{p})$ are transformed to Cartesian coordinates $(m_\text{p},n_\text{p})$. A shift by the horizontal offset candidate $d_i$ (defined within the search range $s$) yields the coordinates $(m_\text{p},n_\text{p}+d_i)$, which are transformed back to polar coordinates $(r_{\text{p},d_i},\phi_{\text{p},d_i})$.
Re-projecting these shifted coordinates back into the fisheye domain by employing:
%\vspace{-0.2cm}
\begin{equation}
\label{eq:fw}
r_{\text{f},d_i} = 2f\sin\left(\frac{1}{2}\arctan\left(\frac{r_{\text{p},d_i}}{f}\right)\right)\:\: \text{and}\:\:\: \phi_{\text{f},d_i} = \phi_{\text{p},d_i}\: ,
%\vspace{-0.15cm}
\end{equation}
and transforming them to a Cartesian representation then yields the coordinates with which to extract the luminance values to be compared against the reference block.
For more details on the luminance value extraction from the reference image, the interested reader is referred to~\cite{eichenseer2015motionfish}.
Using the SSD to minimize the error between the estimated block and the reference block finally yields the optimum candidate which is stored as the disparity map entry $d$ for pixel $(m,n)$.
This procedure is repeated for each pixel of the fisheye image.
Please note that the re-projection also only manipulates the pixel coordinates.
There are no actual distortion corrected images involved, as this would result in either images of infinite dimensions or a significant loss in terms of FOV.
In~\cite{eichenseer2016motioncalib}, a compensation for ultra wide angles ($\theta > \pi/2$) is described. This strategy is also included here; for details, please refer to the original publication.

By employing our proposed disparity estimation method, it is possible to get more accurate estimation results while still only having to store the horizontal displacement in the disparity map.
In contrast to our previous work, the proposed adaptation for disparity estimation is less costly as it requires no hybridization and is further limited to a one-dimensional search without causing a loss in quality.
In the following, we describe how the obtained disparity map can be used for synthesizing intermediate fisheye views.
\begin{figure}[t]
\small
\vspace{0.1cm}
\centering
\psfrag{a1}[cc][tc]{{\color{brownish}Intermediate view}}
\psfrag{a2}[cc][tc]{{\color{brownish}Right view}}
\psfrag{a5}[cc][tc]{{\color{brownish}Left view}}
\psfrag{a3}[tc][tc]{$(m,n)$}
\psfrag{a4}[tc][tc]{$(m_\frac{d}{2},n_\frac{d}{2})$}
\psfrag{a7}[bc][bc]{$n$}
\psfrag{a8}[br][br]{$m$}
\centerline{\includegraphics[width=\columnwidth]{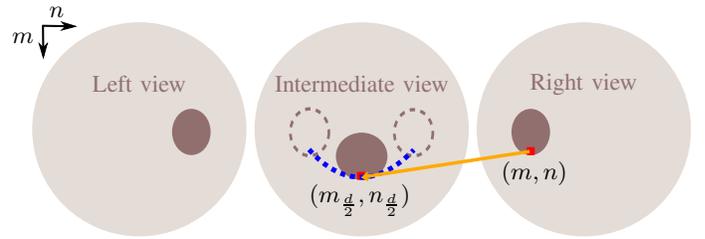}}
\vspace{-0.2cm}
\caption{Intermediate fisheye view synthesis. The shift by $d/2$ is calculated in the perspective domain and translates to a two-dimensional offset (orange) in the fisheye domain.}
\vspace{-0.5cm}
\label{fig:proposed}
\end{figure}

%\vspace{0.4cm}
%\textit{Intermediate Fisheye View Synthesis}
%\vspace{0.2cm}
\subsection*{Intermediate Fisheye View Synthesis}

%Disparity estimation is fundamental when creating new views from a given camera setup.
%With intermediate view synthesis, arbitrary new views between rectified available views can be created.
%Applications include free-viewpoint television, for example, where the user can interactively choose and change views.
%Since it is not possible to capture a scene with an infinite number of cameras so as to provide views from any conceivable view point, new views have to be generated from a limited number of existing ones.
%In the most simple case, an intermediate view is to be created from a stereo camera setup.

To save storage space, it is sufficient to store the perspective horizontal offset between the left fisheye view and the right fisheye view for each pixel.
As the actual offset is expressed in both the horizontal and the vertical dimension, the projections (\ref{eq:bw}) and (\ref{eq:fw}) between the fisheye and the perspective domain must again be employed so as to re-obtain the actual two-dimensional fisheye offset.
In order to generate an intermediate fisheye view, the dense disparity map previously calculated can be used as follows.

For each fisheye pixel position $(m,n)$, the polar coordinates $(r_\text{f},\phi_\text{f})$ are obtained and projected to the perspective domain according to (\ref{eq:bw}), where we can then shift the perspective pixel coordinates $(m_\text{p},n_\text{p})$ by half the disparity, i.\,e., $d/2$.
Just like before, $d$ describes the disparity map entry for positon $(m,n)$.
After the re-projection of the shifted perspective coordinates to the fisheye domain according to (\ref{eq:fw}), we obtain the polar fisheye coordinates $(r_{\text{f},d/2},\phi_{\text{f},d/2})$, which are further transformed into Cartesian coordinates $(m_{d/2},n_{d/2})$. That way, the actual target positions for warping the luminance values are obtained and can be used for the view synthesis:
%\vspace{-0.15cm}
\begin{equation}
%\label{eq:ivs}
I_\text{intermediate}(m_\frac{d}{2},n_\frac{d}{2}) = I_\text{right}(m,n)\: .
%\vspace{-0.05cm}
\end{equation}
Fig.~\ref{fig:proposed} provides a visualization that illustrates how an object changes shape and position between views and how a horizontal offset between perspective views translates to a two-dimensional offset between fisheye views.
An equivalent representation of the perspective disparity is given in blue.
In the following, we evaluate our proposed disparity estimation method in the context of intermediate view synthesis for synthetically generated and real-world fisheye images.

%\vspace{0.2cm}
\section{Simulation Setup and Results}
\label{sec:simulation}
%\vspace{0.1cm}
%not yet calibrated! potential for further small gains
%no consideration of depth/occlusions in both methods

\begin{table}[h]
\small
%\vspace{0.05cm}
\caption{Test sequences and frames used.}
\label{tab:setup}
\vspace{-0.2cm}
\centering
\renewcommand\arraystretch{0.9}
\begin{tabularx}{\columnwidth}{p{0.35cm}p{1.5cm}cc}
\toprule
&\textbf{Sequence} & \textbf{Frame numbers}  & \textbf{Frame offsets to right}\\
& & \textbf{of left views} & \textbf{views (base lines)}\\
\midrule
\multirow{6}{2cm}{\rotatebox[]{90}{Synthetic}}&\textit{Clips} & 10--30 & 2, 4, 6 \\
&\textit{Pencils} & 20--40 & 2, 4, 6 \\
&\textit{Street} & 400--420 & -2, -4, -6 \\
&\textit{PoolA} & 30--50 & 2, 4, 6 \\
&\textit{LivingroomC} & 20--40 & 10, 20, 30 \\
&\textit{HallwayD} & 40--60 & 2, 10, 20 \\
\addlinespace
\multirow{6}{1cm}{\rotatebox[]{90}{Real-World}}&\textit{LibraryB} & 100--120 & 2, 10, 20\\
&\textit{ClutterA} & 100--120 & 40, 70, 100 \\ 
&\textit{ClutterB} & 100--120 & 40, 70, 100 \\ 
&\textit{LectureB} & 100--120 &  -10, -20, -30 \\
&\textit{DriveD} & 400--420 & -2, -6, -10 \\
&\textit{DriveE} & 130--150 & -2, -4, -6 \\
\bottomrule
\end{tabularx}
\vspace{-0.5cm}
\end{table}

\begin{table*}[p]
\small
%\vspace{0.05cm}
\caption{Intermediate view synthesis results (luminance PSNR in dB) averaged over the number of views that were generated per sequence. Results are given for three frame offsets (cf. Table~\ref{tab:setup}), denoted as small, medium, and large base line.}
\label{tab:psnr}
\vspace{-0.15cm}
\centering
\renewcommand\arraystretch{0.95}
\begin{tabularx}{\textwidth}{p{1.28cm}cc|ccc|ccc|ccc}
\toprule
& & & \multicolumn{3}{c}{Small Base Line} & \multicolumn{3}{c}{Medium Base Line} & \multicolumn{3}{c}{Large Base Line}\\
%\addlinespace
\textbf{Sequence} & \hspace{-0.05cm}\textbf{\#Views}\hspace{-0.15cm} & $w$/$s$ & \textbf{BM-IVS} & \textbf{Fisheye-IVS} & $\Delta$ & \textbf{BM-IVS} & \textbf{Fisheye-IVS} & $\Delta$ & \textbf{BM-IVS} & \textbf{Fisheye-IVS} & $\Delta$ \\
\midrule
\textit{Clips} & 21 & 8/256 & 20.22 & 28.93 & 8.71 & 15.28 & 22.44 & 7.16 & 13.65 & 19.89 & 6.24 \\
\textit{Pencils} & 21 & 8/256 & 34.64 & 37.25 & 2.61 & 29.47 & 31.34 & 1.87 & 26.73 & 28.05 & 1.32 \\
%Street & 21 & 8/256 & 21.54 & 26.23 & 4.70 & 20.36 & 23.24 & 2.88 & 19.28 & 21.47 & 2.18 \\
\textit{Street} & 21 & 8/256 & 21.57 & 26.28 & 4.71 & 20.39 & 23.33 & 2.94 & 19.35 & 21.78 & 2.42 \\
\textit{PoolA} & 21 & 8/256 & 34.83 & 36.74 & 1.91 & 30.51 & 32.00 & 1.49 & 28.48 & 29.33 & 0.86 \\
%PoolNightA & 4 & 8/256 &  &  &  & & & & & & \\
\textit{LivingroomC} & 21 & 8/256 & 32.71 & 37.49 & 4.78 & 27.73 & 32.99 & 5.26 & 25.36 & 29.85 & 4.49 \\
\textit{HallwayD} & 21 & 8/256 & 34.45 & 36.76 & 2.31 & 26.19 & 31.85 & 5.67 & 23.11 & 28.83 & 5.73 \\
%HallwayD & 4 & 8/64 & 32.15 & 37.07 & 4.91 & 26.67 & 30.74 & 4.06 & 23.82 & 24.66 & 0.83 \\
\addlinespace
\textit{LibraryB} & 21 & 8/256 & 38.14 & 39.19 & 1.05 & 28.35 & 31.78 & 3.42 & 23.63 & 26.94 & 3.30\\
\textit{ClutterA} & 21 & 8/256 & 32.96 & 33.71 & 0.75 & 31.40 & 33.36 & 1.97 & 27.12 & 29.17 & 2.05\\
\textit{ClutterB} & 21 & 8/256 & 31.68 & 32.53 & 0.85 & 28.75 & 30.01 & 1.26 & 25.52 & 28.29 & 2.77\\
%LectureB & 21 & 8/256 & 27.63 & 29.94 & 2.30 & 22.73 & 26.02 & 3.29 & 20.53 & 23.78 & 3.24\\
\textit{LectureB} & 21 & 8/256 & 27.90 & 29.96 & 2.06 & 24.38 & 27.80 & 3.42 & 22.55 & 26.09 & 3.54\\
\textit{DriveD} & 21 & 8/256 & 31.52 & 32.94 & 1.43 & 27.53 & 29.33 & 1.79 & 25.35 & 27.23 & 1.89\\
\textit{DriveE} & 21 & 8/256 & 27.62 & 29.14 & 1.52 & 24.33 & 26.71 & 2.38 & 22.27 & 23.26 & 0.99\\
%\midrule
%\multicolumn{6}{r}{\footnotesize T: Translation, Z: Zoom, P: Pan, R: Rotation} \\
\bottomrule
\end{tabularx}
%\vspace{0.1cm}
\end{table*}

To evaluate our proposed disparity estimation and view synthesis methods, six synthetically generated as well as six actually captured real-world sequences with an FOV of 185 degrees are used.
All sequences are part of the publicly available fisheye data set introduced in \cite{eichenseer2016dataset}.
Since this data set was captured with a single camera, we simulate a stereo setup by including only sequences with a purely horizontal camera motion in the test set.
As the camera motion is uniform in the considered synthetic sequences, left and right views can be selected and ground truth intermediate views are immediately available for evaluation purposes.
For the real-world sequences, the camera motion may have been less stable, thus providing slightly less accurate ground truth images, but we will still show that the proposed method outperforms the reference.
Table~\ref{tab:setup} summarizes the sequences that are part of the test set.
The top six sequences are synthetic, the bottom six were captured with a fisheye camera, details on which can be found in~\cite{eichenseer2016dataset}.
The frame number is given for each left view that was included in the test.
The right view is obtained by adding the given frame offset to the frame number of the left view.
Please note that a positive frame offset is given for all sequences that exhibit a left-to-right camera motion. In contrast, a negative frame offset is given for those sequences that feature a camera motion from right to left.
%Please note that a negative frame offset is given for sequences \textit{Street}, \textit{LectureB}, \textit{DriveD}, and \textit{DriveE} since the camera moves from right to left instead of left to right in those sequences.
For each sequence, 21 view pairs are tested. Three frame offsets are used per each pair, simulating a small, medium, and large base line between the two cameras.
In accordance with \cite{eichenseer2015motionfish} and \cite{eichenseer2016motioncalib}, we use SSD as a dissimilarity metric; if desired, SSD can easily be substituted by another metric, though.
For the intermediate view synthesis, we synthesize the exact middle frame between the left view and the right view.
We use Delaunay triangulation followed by cubic interpolation to obtain integer target positions both in the conventional block matching method and in the proposed approach.
This is mainly to avoid disadvantageous rounding operations as $(m_{d/2},n_{d/2})$ may describe non-integer numbers, but it simultaneously serves as the hole-filling algorithm.
In this work, we do not yet handle occlusions and disocclusions in either the reference or the proposed method.
If a certain integer target position is written to more than once, the last warped value is kept.

\begin{figure*}[p]
%\small
%\vspace{0.1cm}
\centering
\psfrag{a1}[cB][cB]{\textbf{Left view}}
\psfrag{a2}[cB][cB]{\textbf{Right view}}
\psfrag{a3}[cB][cB]{\textbf{Intermediate view}}
\psfrag{a4}[cB][cB]{\textbf{BM-IVS}}
\psfrag{a5}[cB][cB]{\textbf{Fisheye-IVS}}
\centerline{\includegraphics[width=\textwidth]{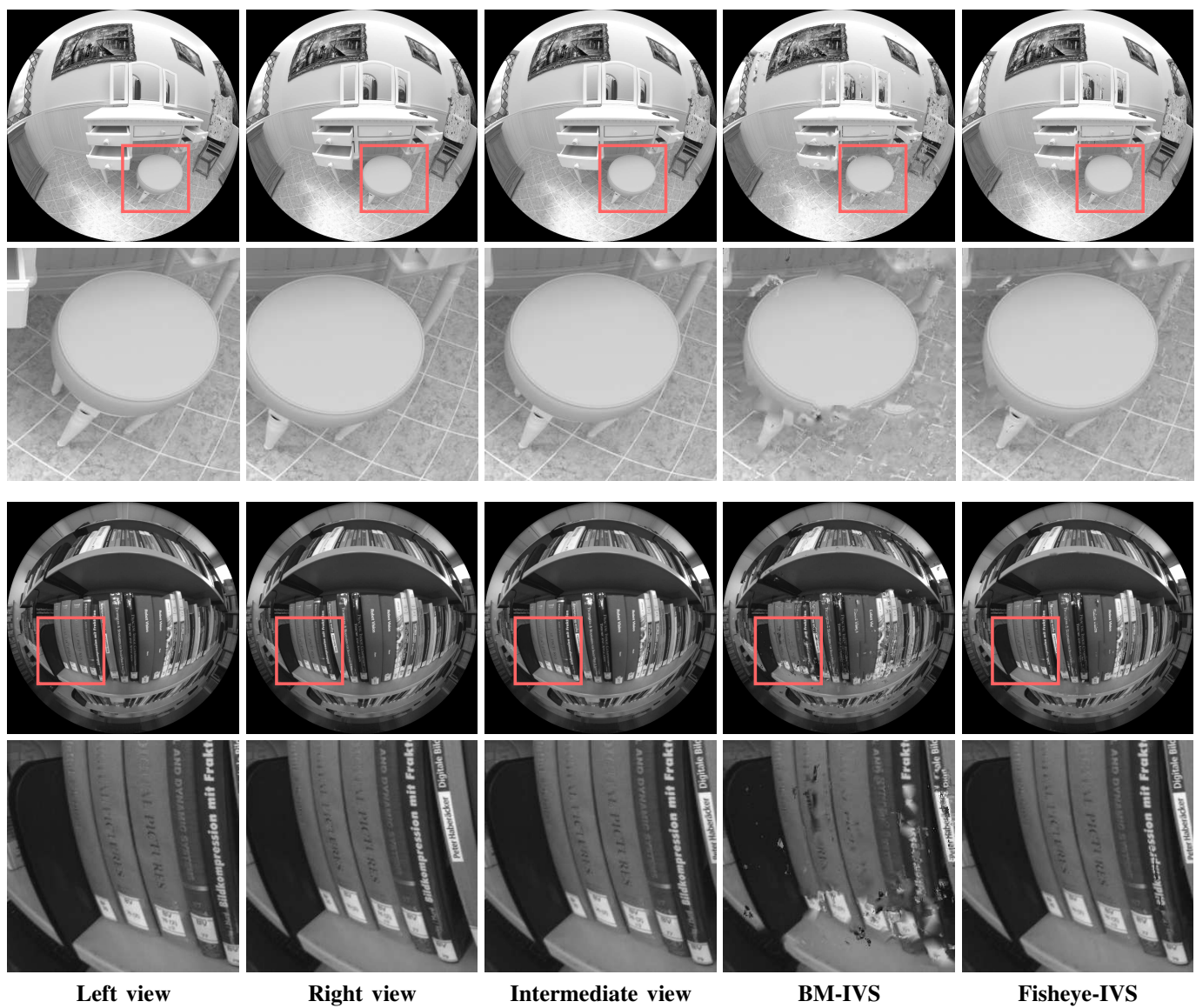}}
%\vspace{-0.15cm}
\caption{Visual examples for the intermediate views as obtained by BM-IVS and Fisheye-IVS. First row: \textit{HallwayD}, frames 50 (left view), 70 (right view), and 60 (ground truth intermediate view). Third row: \textit{LibraryB}, frames 100 (left view), 120 (right view), and 110 (ground truth intermediate view). Second and fourth row: detail examples of the images above.}
\label{fig:visualres}
%\vspace{-0.1cm}
\end{figure*}

The conventional intermediate view synthesis based on disparity estimation via block matching is referred to as BM-IVS.
Our proposed method is called Fisheye-IVS.
For both methods, we compared the synthesized intermediate views against their ground truth views and used the luminance PSNR to objectively evaluate the quality of the result. The PSNR was computed excluding the black peripheral areas beyond the 185 degree border, i.\,e., those sensor regions not hit by light.
Table~\ref{tab:psnr} summarizes the PSNR results for the entire test set consisting of the twelve sequences and three base line settings.
Blocks of $17\times 17$ pixels were used, so that a support margin of $w = 8$ pixels was defined around each pixel position.
The one-dimensional search range was set to $s=256$, so that horizontal candidate offsets in the range $[0, 256]$ were tested for each block.
Evidently, the proposed Fisheye-IVS outperforms BM-IVS in all instances and achieves significant gains without adding a second dimension to the disparity information.
PSNR values decrease for larger base lines, as the search range may be too short to accurately capture the disparity and occlusions are more pronounced for the wider spacings. 
Sequences with a fast camera motion (particularly \textit{DriveD} and \textit{DriveE} which were captured out of a driving car) automatically result in rather large base lines.
To substantiate the objective results, Fig.~\ref{fig:visualres} provides representative visual examples for the large base line settings of \textit{HallwayD} and \textit{LibraryB}.
%Similar visual results are obtained for all sequences tested.
%The visual example clearly confirms the objective results and shows that a fisheye adaptation is indeed worthwhile.
For a real stereo setup that uses two distinct cameras, additional factors such as diverging internal camera settings and brightness inconstancies may also have to be accounted for when performing an intermediate view synthesis.
This does not directly or exclusively affect the fisheye adaption, however, so that results similar to those shown in this paper may be expected.

%We expect similar results for a real two-camera setup, albeit with decreased absolute PSNR values for both methods due to differing recording conditions such as diverging internal camera settings.

%\vfill\pagebreak
%\vspace{-0.2cm}
\section{Conclusion}
\label{sec:conclusion}
%\vspace{-0.1cm}
In this paper, we analyzed the effects of a conventional disparity estimation method based on horizontal block matching on fisheye images.
As the inherent radial distortion of fisheye images causes a significant deterioration of the estimation results, we proposed a novel adaptation that exploits knowledge about the fisheye projection function. 
Using a fisheye-to-perspective projection, the pixel coordinates are transformed to their perspective representation, for which a horizontal search is able to find better matches and thus create a more accurate disparity map. In doing so, no actual distortion correction of the fisheye images had to be performed and neither was it necessary to provide two-dimensional disparity information as would otherwise be needed for fisheye views.
In contrast to our previous work, both hybridization and searching in two dimensions were avoided. 
Intermediate view synthesis was employed on both synthetic and real-world fisheye images to show the benefits of our adapted approach.
Significant luminance PSNR gains were achieved for the entire test set and further reflected by means of visual examples.
Our proposed disparity estimation method can be easily adapted to any kind of distortion or deviation from the pinhole model, provided the projection function is known; furthermore, other dissimilarity measures may be easily incorporated.
Current research includes analyses of view synthesis based on more than one reference view, the inclusion of calibration information to improve the results for the real-world sequences, occlusion-handling, and adapting our scheme towards more sophisticated approaches. %\cite{sunwoo}
Further points of interest include depth information in the context of fisheye data as well as multi-view coding and frame rate up-conversion for fisheye sequences.

%\vfill
%\pagebreak

%\vspace{0.2cm}
\section*{Acknowledgment}
%\vspace{0.15cm}

This work was supported by the Research Training Group 1773 “Heterogeneous Image Systems”, funded by the German Research Foundation (DFG).
%\vspace{0.3cm}

\bibliographystyle{IEEEtran}
\bibliography{refs}

\end{document}